\documentclass{aa}
\usepackage{graphicx}
\usepackage{natbib}
\bibpunct{(}{)}{;}{a}{}{,}
\begin{document}
\title{The fundamental definition of `radial velocity'}


\author{Lennart Lindegren
        \and
        Dainis Dravins
        }

\offprints{L.~Lindegren}

\institute{Lund Observatory, Box 43, SE-22100 Lund, Sweden\\
           \email{lennart@astro.lu.se, dainis@astro.lu.se}
          }

\date{Received September 4, 2002; accepted January 31, 2003}

\abstract{
Accuracy levels of metres per second require the fundamental concept
of `radial velocity' for stars and other distant objects to be examined,
both as a physical velocity, and as measured by spectroscopic and
astrometric techniques.  Already in a classical (non-relativistic)
framework the line-of-sight velocity component is an ambiguous concept,
depending on whether, e.g., the time of light emission (at the object)
or that of light detection (by the observer) is used for recording the
time coordinate.  Relativistic velocity effects and spectroscopic
measurements made inside gravitational fields add further complications,
causing wavelength shifts to depend, e.g., on the transverse velocity
of the object and the gravitational potential at the source.
Aiming at definitions that are unambiguous at accuracy levels of
1~m~s$^{-1}$, we analyse different concepts of radial velocity
and their interrelations. At this accuracy level, a strict separation
must be made between the purely geometric concepts on one hand, and
the spectroscopic measurement on the other. Among the geometric
concepts we define \emph{kinematic radial velocity}, which corresponds
most closely to the `textbook definition' of radial velocity as the
line-of-sight component of space velocity; and \emph{astrometric
radial velocity}, which can be derived from astrometric observations.
Consistent with these definitions, we propose strict definitions also of
the complementary kinematic and astrometric quantities, namely transverse
velocity and proper motion. The kinematic and astrometric radial
velocities depend on the chosen spacetime metric, and are accurately
related by simple coordinate transformations.  On the other hand, the
observational quantity that should result from accurate spectroscopic
measurements is the \emph{barycentric radial-velocity measure}.
This is independent of the metric, and to first order
equals the line-of-sight velocity.  However, it is not a physical
velocity, and cannot be accurately transformed to a kinematic or
astrometric radial velocity without additional assumptions and
data in modelling the process of light emission from the source,
the transmission of the signal through space, and its recording by
the observer.  For historic and practical reasons, the spectroscopic
radial-velocity measure is expressed in velocity units as
$cz_{\rm B}$, where $c$ is the speed of light and $z_{\rm B}$ is
the observed relative wavelength shift reduced to the solar-system
barycentre, at an epoch equal to the barycentric time of light
arrival. The barycentric radial-velocity measure and the astrometric
radial velocity are defined by recent resolutions adopted by the
International Astronomical Union (IAU), the motives and
consequences of which are explained in this paper.
}

\maketitle

\section{The need for stringent definitions}
\label{sec:intro}

\emph{Radial velocity} is an omnipresent concept in astronomy, and
a quantity whose precision of determination has improved
significantly in recent years.  Its meaning is generally understood
as the object's motion along the line of sight, a quantity normally
deduced from observed spectral-line displacements, interpreted as
Doppler shifts.  However, despite its ubiquity, there has not
existed any physically stringent definition of `radial velocity'
with an accuracy to match currently attainable measuring precisions.
Two first such definitions -- one for the result of spectroscopic
observations, and one for the geometric (astrometric) concept of
radial velocity -- were adopted at the General Assembly of the
International Astronomical Union (IAU), held
in 2000.  The purpose of this paper is to explain their background,
the need for such definitions, and to elaborate on their
consequences for future work.  Thus, the paper is \emph{not} about
the detailed interpretation of observed spectral-line displacements
in terms of radial motion, nor about the actual techniques for making
such measurements; rather, it is the definition of `radial velocity'
itself, as a geometric and spectroscopic concept, that is discussed.

The need for a strict definition has become urgent in recent years
as a consequence of important developments in the techniques for
measuring stellar radial velocities, as well as the improved
understanding of the many effects that complicate their
interpretation. We note in particular the following circumstances:

\emph{Precision and accuracy of spectroscopic measurements:}
Spectro\-scopic measurement precisions 
are now reaching (and surpassing) levels of metres per second.
In some applications, such as the search for (short-period) 
extrasolar planets or stellar oscillations, it may be sufficient to
obtain differential measurements of wavelength shifts, in which 
case internal precision is adequate and there is no need for an 
accurate definition of the zero point. However, other applications 
might require the combination of data from different observatories, 
recorded over extended periods of time, and thus the use
of a common reference point. 
Examples could be the study of long-term variations due to stellar 
activity cycles and searches for long-period 
stellar companions. 
Such applications call for data that are not only precise, but also 
accurate, i.e., referring to some `absolute' scale of measurements.
However, the transfer of high-precision measurements to absolute values 
was previously not possible, partly because there has been no agreement 
on how to make such a transfer, or even on which physical quantity to 
transfer.

In the past, a classical accuracy achieved for measuring stellar
radial velocities has been perhaps 1~km~s$^{-1}$, at which level
most of these issues did not arise, or could be solved by the simple
use of `standard stars'. With current methods and instrumentation, 
the accuracy by which measured stellar wavelengths can be related 
to absolute numbers is largely set by the laboratory sources used 
for spectrometer calibration (lines from iodine cells, lasers, etc.). 
An accuracy level of about 10~m~s$^{-1}$ now seems reachable. Since 
any fundamental definition should be at least some order of magnitude 
better than current performance, the accuracy goal for the definition 
was set to 1~m~s$^{-1}$. This
necessitates a stringent treatment of the radial-velocity concept.

\emph{Ambiguity of classical concepts:}
A closer inspection even of the classical (non-relativistic)
concepts of radial velocity reveals that these are ambiguous at
second order in velocity relative to the speed of light.
For instance, if radial velocity is
defined as the rate of change in distance, one may ask whether
the derivative should be with respect to the time of light
emission at the object, or of light reception at the observer.
Depending on such conventions, differences exceeding
1~m~s$^{-1}$ would be found already for normal stellar
velocities.

\emph{Intrinsic stellar spectroscopic effects:}
On accuracy levels below $\sim$\,1~km~s$^{-1}$, spectral lines in
stars and other objects are generally asymmetric and shifted in
wavelength relative to the positions expected from a
Doppler shift caused by the motion of their centres-of-mass.
Such effects are caused e.g.\ by convective motions in the
stellar atmosphere, gravitational redshift, pressure shifts,
and asymmetric emission and/or absorption components. As a
consequence, the measured wavelengths do not correspond to the
precise centre-of-mass motion of the star.

\emph{Relation between Doppler shift and velocity:}
Even if we agree to express the observed wavelength shift
(whatever its origin) as a velocity, it is not obvious how
that transformation should be made. Should it use the classical
formula $v_r=cz$ (where $c$ is the speed of light and
$z=(\lambda_{\rm obs}-\lambda_{\rm lab})/\lambda_{\rm lab}$
the dimensionless spectral shift),
or the relativistic version (in which case the transverse
velocity must either be known or assumed to be negligible)?
Differences are of second order in $z$,
thus exceeding 1~m~s$^{-1}$ already for `normal' stellar
velocities ($\ga 20$~km~s$^{-1}$), and 100~m~s$^{-1}$ for
more extreme galactic velocities ($\ga 200$~km~s$^{-1}$).

\emph{The role of standard stars:}
Practical radial-velocity measurements have traditionally relied
on the use of standard stars to define the zero point of the
velocity scale. While these have aimed at accuracies of the order
100~m~s$^{-1}$, it has in reality been difficult to achieve
consistency even at this level due to poorly understood systematic
differences depending on spectral type, stellar rotation,
instrumental resolution, correlation masks used, and so on.
Standard stars will probably continue to play a role as a
practical way of eliminating, to first order, such differences
in radial-velocity surveys aiming at moderate accuracy.
However, their relation to high-accuracy measurements needs to
be clearly defined.

\emph{Gravitational redshifts:}
The gravitational potential at the stellar surface causes
all escaping photons to be redshifted by an amount that
varies from $\sim\,$30~m~s$^{-1}$ for supergiants,
$\sim\,$30~km~s$^{-1}$ for white dwarfs, and much greater values
for neutron stars and other compact objects. Even for a given
star, the precise shift varies depending on the height at
which the spectral lines are formed. The observed shift
also depends on the gravitational potential at the observer,
and therefore on the observer's distance from the Sun.

\emph{Astrometric determination of radial motion:}
Current and expected advances in astrometry enable the accurate
determination of stellar radial motions without using spectroscopy
\citep{drav+99a}, i.e., based on purely geometric measurements
such as the secular change in trigonometric parallax. Comparison
of such velocities with spectroscopic measurements could obviously
give a handle on the intrinsic stellar effects mentioned above,
but how should such a comparison be made? How does the astrometric
radial velocity differ conceptually from the spectroscopically
determined velocity?

\emph{Accurate reference systems for celestial mechanics and
astrometry:} The rapid development of observational accuracies
in astrometry and related disciplines has made it necessary to
introduce new conventions and reference systems, consistent
with general relativity at sub-microarcsecond levels
\citep{john+00}. Radial velocity, regarded as a component of
space velocity, obviously needs to be considered within the
same framework.

\emph{Cosmology:} Ultimately, spectroscopic measurements of distant
stars are also affected by cosmological redshift.
To what extent does also the local space to nearby stars take part
in the general expansion of the Universe?  What is an actual
`velocity', and what is a change of spatial coordinates?  Since
such factors are generally not known to the spectroscopic observer,
it is impossible to convert the observed shift into a precise
kinematic quantity.

\begin{figure}
\centering
\resizebox{\hsize}{!}{\includegraphics{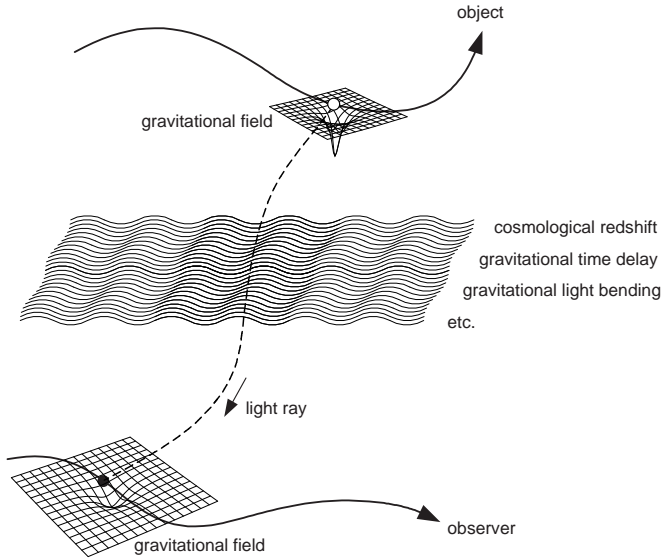}}
\caption{Whether a radial velocity is measured spectroscopically
or geometrically, the relevant astronomical event consists of
several parts: the motion of the observed object; the emission
of an electromagnetic signal from the object, symbolised by the
open circle; the propagation of the signal through space; the
motion of the observer; and the reception of the signal by the observer,
symbolised by the black dot \citep[figure partially adapted from][]{klio00}.
}
\label{fig:1}
\end{figure}

From these examples it should be clear that the naive concept
of radial velocity, as the line-of-sight component of the stellar 
velocity vector measured by the Doppler shift of the spectral lines, 
is far too simplistic when aiming at accuracies 
much better than 1~km~s$^{-1}$. 
To arrive at a consistent set of definitions applicable to the various
classes of observations, it is necessary to consider all the
phases of an astronomical event (Fig.~\ref{fig:1}).
These include the motion of the star; the emission of a light
signal from the star and its propagation to the observer through
varying gravitational fields and possibly expanding space; the
motion of the observer; and the reception and measurement of the
signal by the observer.%
\footnote{For conciseness, we will from here on use the word
`star' to denote any observed object far outside the solar
system, and `light' to denote electromagnetic radiation from
that object.}
The observer has detailed knowledge only of the last two phases
of the event (symbolised by the black dot in Fig.~\ref{fig:1}),
while the interpretation of the previous phases requires additional
assumptions or modelling. The result of a measurement should
ideally be specified in a way that is neutral with respect to
such interpretation. As we shall see (Sect.~\ref{sec:rvm}), this
leads to the definition of \emph{barycentric radial-velocity
measure} as the desired result of a spectroscopic measurement. However,
to relate this quantity to physical motion in even simple
situations, a model is required which incorporates all phases of
the event as illustrated in the figure. This, in turn, leads to the
introduction of additional
geometric quantities, viz.\ \emph{kinematic radial velocity}
and \emph{astrometric radial velocity} (Sect.~\ref{sec:math}).

The remainder of this paper is organised as follows.
Section~\ref{sec:whatis} contains a preliminary heuristic discussion
of the radial-velocity concept as such; the purpose is to point
out the inadequacy and ambiguity of standard notions, without
providing a solution.
Section~\ref{sec:spec} then gives an overview of spectroscopic
radial-velocity determinations, highlighting effects other than
trivial stellar motion that influence the outcome of such
measurements.
The accurate meaning of the geometric and spectroscopic quantities
is evaluated in the more technical Sections~\ref{sec:math} and
\ref{sec:rvm}, leading up to the IAU resolutions,
whose practical implications are considered in Sect.~\ref{sec:iau}.
Some unsolved issues, beyond the scope of the present definitions,
are briefly discussed in Sect.~\ref{sec:unsolved}.
The Appendix contains the full text of the two IAU resolutions.

\section{What is meant by `radial velocity'?}
\label{sec:whatis}

In this Section we examine the common notions of `radial velocity'
from a somewhat naive viewpoint, in order to highlight some of the
difficulties associated with this apparently simple concept.
Additional complications in the interpretation of spectroscopic
shifts are discussed in Sect.~\ref{sec:spec}.

\subsection{Geometric concepts}
\label{sec:geom}

The \emph{Encyclopedia of Astronomy and Astrophysics} defines
the radial velocity of a star as `the component of its motion
along the line of sight of the observer' \citep{lath01}. The
\emph{Explanatory Supplement to the Astronomical Almanac}
gives an alternative definition, `the rate of change
of the distance to an object' \citep{seid92}. Both agree with
common notions about radial velocity, but are they equivalent?
Let us start by examining this question in a purely classical
framework, without the complications of relativity, but taking
into account the finite speed of light ($c$).

In a Euclidean metric with origin at the solar-system barycentre
and with $t$ denoting coordinate time, let $\vec{r}_\ast(t)$
be the motion of the star,
$\vec{v}_\ast = {\rm d}\vec{r}_\ast/{\rm d}t$
its barycentric space velocity,
$r_\ast = \left| \vec{r}_\ast \right|$ the barycentric distance,
and $\vec{u} = \vec{r}_\ast/r_\ast$
the barycentric direction to the star. Following the first of the
two definitions quoted above, the radial velocity $v_r$ is
the component of $\vec{v}_\ast$ along $\vec{u}$, or
\begin{equation}\label{eq:01}
  v_r = \vec{u} ' \vec{v}_\ast \, ,
\end{equation}
where the prime ($'$) denotes scalar multiplication of vectors.
From $\vec{r}_\ast=\vec{u}r_\ast$ we can write the space velocity
\begin{equation}\label{eq:02}
  \vec{v}_\ast \equiv {\displaystyle\frac{{\rm d}\vec{r}_\ast}{{\rm d}t}}
  = {\displaystyle\frac{{\rm d}\vec{u}}{{\rm d}t}}\,r_\ast
  + \vec{u}\,{\displaystyle\frac{{\rm d}r_\ast}{{\rm d}t}} \, .
\end{equation}
Taking the scalar product with $\vec{u}$ we have, since $\vec{u}'\vec{u}=1$
and $\vec{u}'({\rm d}\vec{u}/{\rm d}t)=
\frac{1}{2}{\rm d}(\vec{u}'\vec{u})/{\rm d}t = 0$,
\begin{equation}\label{eq:03}
  \vec{u} ' \vec{v}_\ast = {\displaystyle\frac{{\rm d}r_\ast}{{\rm d}t}}\, .
\end{equation}
The right member apparently corresponds to the second definition
quoted above. Comparing Eqs.~(\ref{eq:01}) and (\ref{eq:03}) it therefore
appears that we have proved the equivalence of the two definitions.

However, the situation is more complex when the finite speed of light
is considered. The observation involves (at least) two different
times, viz.\ the time of light emission at the star ($t_\ast$) and
the time of light reception at the observer ($t_{\rm obs}$),
cf.\ Fig.~\ref{fig:1}. The second definition, `the rate of change of
the distance', is in fact ambiguous because it is not specified which
time is used to compute the time derivative.
Clearly the $t$ in Eq.~(\ref{eq:03}) 
must be the same as used in describing
the motion of the star, $\vec{r}_\ast(t)$, which 
should be independent
of the observer and therefore corresponding to the time of light emission.
However, when describing an observed phenomenon, such as a measurement
of the line-of-sight velocity of a star, it is more natural to refer it
to the time of light reception $t_{\rm obs}$.%
\footnote{There is an analogous problem in the definition of
proper motion, i.e.\ the rate of change in direction
$\vec{u}$, but the consensus is that proper motion
means ${\rm d}\vec{u}/{\rm d}t_{\rm obs}$, not
${\rm d}\vec{u}/{\rm d}t_\ast$; cf.\ Sect.~\ref{sec:astr}.}

The two instants $t_\ast$ and $t_{\rm obs}$ are related by the light-time
equation, which for an observer at the solar-system barycentre (and
ignoring gravitational time delay) is simply
\begin{equation}\label{eq:04}
  r_\ast = c(t_{\rm obs} - t_\ast) \, .
\end{equation}
Depending on which $t$ is used to compute the `rate of change' in the
second definition, we obtain by means of the light-time equation two
different expressions for $v_r$:
\begin{equation}\label{eq:05}
  \left.
    \begin{array}{lllll}
       v_r' &\equiv&
         {\displaystyle\frac{{\rm d}r_\ast}{{\rm d}t_\ast}}
         &=& c\left({\displaystyle\frac{{\rm d}t_{\rm obs}}
           {{\rm d}t_\ast}}-1\right) \\[16pt]
       v_r'' &\equiv&
         {\displaystyle\frac{{\rm d}r_\ast}{{\rm d}t_{\rm obs}}}
         &=& c\left(1-{\displaystyle\frac{{\rm d}t_\ast}
           {{\rm d}t_{\rm obs}}}\right)
    \end{array}
  \quad\right\}
\end{equation}
The difference, $v_r'-v_r'' =
v_r' v_r''/c\simeq v_r^2/c$, exceeds 1~m~s$^{-1}$
for $|v_r|\ga 20$~km~s$^{-1}$ and 100~m~s$^{-1}$ for
$|v_r|\ga 200$~km~s$^{-1}$. Since relative velocities in our Galaxy
are typically tens of km~s$^{-1}$, and may reach several hundred
km~s$^{-1}$, the ambiguity has practical relevance in the context of
precise stellar radial velocities.

It is seen from Eq.~(\ref{eq:05}) that the ambiguity arises when the
quantity ${\rm d}t_{\rm obs}/{\rm d}t_\ast$ is transformed into a
velocity, i.e.\ when a model is used to interpret the data.
${\rm d}t_{\rm obs}/{\rm d}t_\ast$, on the other hand, is a direct,
model-independent relation between the basic events of light
emission and reception. From an observational viewpoint, we could
therefore regard the dimensionless quantity
${\rm d}t_{\rm obs}/{\rm d}t_\ast$ as more fundamental than either
$v_r'$ or $v_r''$.

\subsection{Doppler shift}
\label{sec:dopp}

The result of a spectroscopic line-shift measurement may be
expressed by means of the dimensionless redshift variable
\begin{equation}\label{eq:z}
  z = \frac{\lambda_{\rm obs} - \lambda_0}{\lambda_0}
    = \frac{\nu_0 - \nu_{\rm obs}}{\nu_{\rm obs}} \, ,
\end{equation}
where $\lambda_0$ ($\nu_0$) is the rest-frame wavelength (frequency).
The redshift is often converted to a conventional velocity
using some standard formula, the simplest being
\begin{equation}\label{eq:cz1}
  v_r^{(1)} = c\,\frac{\lambda_{\rm obs} - \lambda_0}{\lambda_0}
            = cz \, .
\end{equation}
However, an alternative conversion is obtained by considering the
relative shift in frequency rather than in wavelength, viz.:
\begin{equation}\label{eq:cz2}
  v_r^{(2)} = c\,\frac{\nu_0-\nu_{\rm obs}}{\nu_0}
            = \frac{cz}{1+z} \, .
\end{equation}
This last expression has traditionally been used in radio
astronomy \citep[e.g.][]{walk87}, although the practice is
discouraged due to the risk of confusion with the earlier
expression \citep{cont+74,mull+77}.
$v_r^{(1)}$ is sometimes called the `optical velocity' and
$v_r^{(2)}$ the `radio velocity' \citep{grei+02}.

For use with large velocities, the following formula is often
recommended \citep[e.g.][]{lang74}:
\begin{equation}\label{eq:cz3}
  v_r^{(3)} = c\,\frac{(1+z)^2-1}{(1+z)^2+1} \, .
\end{equation}
The expression is derived from the special-relativistic
Doppler formula by assuming purely radial motion (cf.\ below).

Thus we have at least three different conventions for transforming
$z$ into $v_r$ that are more or less `standard' in astronomy.
From the series expansions
\begin{equation}\label{eq:czexp}
  \left.
    \begin{array}{lll}
      v_r^{(1)} &=& cz \\[3pt]
      v_r^{(2)} &=& c\,(z-z^2+z^3-z^4+\cdots ) \\[3pt]
      v_r^{(3)} &=& c\,(z-\frac{1}{2}z^2+\frac{1}{4}z^4-\cdots )
    \end{array}
  \quad\quad\right\}
\end{equation}
it is seen that the differences between the three conventions
are of second order ($cz^2 \sim v_r^2/c$).

Equation~(\ref{eq:cz3}) ignores the star's transverse velocity ($v_t$).
The complete Doppler formula from special relativity reads
\citep{lang74}:
\begin{equation}\label{eq:zsr}
  1+z=\frac{1+v_r/c}{\sqrt{1-v^2/c^2}} \, ,
\end{equation}
where $v=(v_r^2+v_t^2)^{1/2}$ is the total velocity relative the
observer. Solving for $v_r$ and expanding in powers of $z$ and
$u=v_t/c$ we obtain a fourth expression for $v_r$:
\begin{equation}\label{eq:zsrexp}
  v_r^{(4)} = c\left[ z - {\textstyle\frac{1}{2}}(z^2+u^2)
  + {\textstyle\frac{1}{4}}zu^2
  + {\textstyle\frac{1}{4}}(z^4-u^4) - \cdots
  \right] .
\end{equation}
Comparing with the third variant of Eq.~(\ref{eq:czexp})
it is seen that the transverse
Doppler effect is $\sim v_t^2/2c$, i.e.\ typically of a similar size
as the differences among the expressions in Eq.~(\ref{eq:czexp}).

Thus, various conventions exist for converting the observed Doppler
shift into a radial velocity; the differences are of order $v^2/c$,
exceeding 1~m~s$^{-1}$ for normal galactic stellar velocities and
100~m~s$^{-1}$ for high-velocity stars. Within the framework of
special relativity (thus ignoring the many other effects discussed
below), a `rigorous' transformation from $z$ to $v_r$ is possible,
but only if the transverse velocity of the star is also known.

\subsection{Astrometric determination or radial motion}

Astrometry specialises in measuring the {\em directions} to objects,
and in particular the directional changes caused by the motions of
the object (proper motion) and observer (parallax).  Although such
measurements primarily yield the distances and transverse velocities
of the objects, they are in principle sensitive also to the radial
motion of the objects through various second-order effects. Although
the principles have been known for a long time, it is only with the
high accuracies realised with space techniques that astrometry has
become a practical possibility for radial-velocity measurements.

Different methods exist for the determination of astrometric stellar
radial velocities \citep{drav+99a}.  The most direct method, measuring
the rate of parallax change as a star approaches or recedes, is still
beyond realised accuracies (e.g., even for the nearby high-velocity
Barnard's star, the expected parallax rate is only 34~$\mu$as~yr$^{-1}$),
although it is expected to become measurable in the foreseeable future.
Another method utilises that a star's proper motion changes as a result
of its changing distance from the Sun (`perspective acceleration').
By combining high-accuracy proper motions with measurements of
stellar positions at different epochs, radial-velocity values have
so far been determined for some 20 stars, though only with modest
accuracies \citep[typically a few tens of km~s$^{-1}$;][]{drav+99a}.
However, a third method, applicable to nearby moving clusters such
as the Hyades, whose stars share the same (average) velocity vector,
already permits accuracies on the sub-km~s$^{-1}$ level, and
for several classes of stars has yielded better data than have been
possible to achieve spectroscopically.  Here, parallaxes and proper
motions are combined to determine the apparent secular expansion or
contraction of the angle subtended by a cluster, as it is
approaching or receding.  Using data from the Hipparcos satellite
mission, more than a thousand stars have already been thus studied
\citep{mads+02}, a number to be increased when data from future
astrometry missions become available.

These methods are all based on the same general principle: let $\theta$
be the angular size of an object (the Earth's orbit as seen from the
star, the distance travelled by the star in a given time, or the size
of a stellar cluster) and $r$ its distance; then the assumption is that
$r\theta=$~constant, from which $\dot{r}=-r\dot{\theta}/\theta$. While
the principle is simple enough, the question still remains whether the
derivative ($\dot{\phantom{r}}$) should be taken with respect to the
time of observation, or the time of light emission. Thus, also the
concept of astrometric radial velocities needs a more precise definition.

\section{Limitations of spectroscopic radial-velocity measurements}
\label{sec:spec}

In this Section we highlight the various issues that may limit the
achievable accuracy in stellar radial velocities, as deduced from
spectroscopic observations.

\subsection{Gravitational redshifts}

The gravitational potential induced by a star's mass causes
redshifts of all photons leaving its vicinity.  Across the
Hertzsprung--Russell diagram, the gravitational redshifts change
by three orders of magnitude between white dwarfs (some
30~km~s$^{-1}$) and supergiants (some 30~m~s$^{-1}$).  This
gravitational redshift $v_{\rm grav} = GM/rc$ diminishes with
distance from the stellar centre as $r^{-1}$.  For the Sun,
the value is 636.5~m~s$^{-1}$ for light escaping from the solar
photosphere ($r = R_\odot$) to infinity,%
\footnote{Using $R_\odot=(6.95508\pm 0.00026)\times 10^8$~m
for the solar photospheric radius \citep{brow+98} and
$GM_\odot=1.327124\times 10^{20}$~m$^3$~s$^{-2}$ \citep{stan95}
we get $GM_\odot/Rc=636.486\pm 0.024$~m~s$^{-1}$.}
and 633.5~m~s$^{-1}$ for light intercepted at the Earth's mean
distance from the Sun ($r = 215R_\odot$).
A solar spectral line instead formed at chromospheric heights
(30~Mm, say; $r = 1.04R_\odot$) will have this redshift decreased
by some 20~m~s$^{-1}$, and a coronal line by perhaps 100~m~s$^{-1}$.

For other stars, the shift scales as $(M/M_\odot)(R/R_\odot)^{-1}$,
or as $(g/g_\odot)(R/R_\odot)$, where $g$ is the surface gravity.
Since $R$ and $M$ can rarely be estimated to better than $\sim$\,5\%
for single stars \citep{ande91}, while spectroscopic determinations
of $\log g$ have much larger uncertainties \citep{lebr00}, it is
normally not possible to compute the gravitational redshift to
better than 50~m~s$^{-1}$ for individual single stars.

\subsection{Effects inside stellar atmospheres}

It is well known that photospheric absorption lines in the solar spectrum
are blueshifted by about 400~m~s$^{-1}$ (after correction for the
known gravitational redshift) as a result of convective motions in
the solar atmosphere. In the photospheric granulation, hot (bright)
and rising (blueshifted) convective elements contribute more photons
than the cooler (darker) and sinking (redshifted) gas, thus causing a
net blueshift of the absorption lines \citep{drav82,alle+98a}.
Detailed modelling of stellar atmospheres involving 3-dimensional
and time-dependent hydrodynamics is capable of producing synthetic
spectral lines whose intensity profiles and patterns of wavelength
displacements closely match observations for at least solar-type
stars \citep[e.g.][]{aspl+00,alle+02}.
For main-sequence stars, the predicted convective blueshifts range
between approximately 1000~m~s$^{-1}$ for F-type stars and 200~m~s$^{-1}$
for cooler K-types.

The shift is however not the same for all the lines in a spectrum.
The precise amount of shift depends on the strength of
the absorption line (and hence on the stellar metallicity), since
different lines are formed at different atmospheric depths and thus
experience different granulation contrasts.
For the Sun such (observed and modelled) differential shifts
between differently strong lines in the visual amount to some
200~m~s$^{-1}$, but reach 1000~m~s$^{-1}$ for the hotter F-type
star Procyon \citep{alle+02}.  The shifts further depend on
excitation potential and ionisation level (due to different
conditions of line formation), and on wavelength region (due to
varying granulation contrast).  Actually, in some
wavelength regions, where the lines may originate in atmospheric
layers characterised by convective overshoot (with an inverted
velocity/temperature correlation), the lineshifts may change
sign to become convective redshifts. For the Sun this can be
observed in cores of very strong lines in the optical or
generally in the far ultraviolet \citep{sama91}.

Line profiles are also asymmetric, making the determination of
accurate lineshifts a matter of convention -- where in the line should
the shift be measured? Asymmetries are caused 
not only by convective motions on the stellar surface but also by 
asymmetric emission and/or absorption
components (e.g., due to chromospheric emission or stellar-wind
absorption), by microscopic processes causing wavelength shifts
on the atomic level (e.g., pressure shifts), or macroscopic
circumstances (e.g., gravitational redshift). Further complications
enter for pulsating stars, those with expanding atmospheres, or such
with deviant isotopic compositions.

Since many physical effects thus contribute to the observed
wavelength shifts \citep[e.g.][]{drav02}, it is
not possible to deduce an accurate centre-of-mass motion simply
from the observed differences between wavelengths in the source
and those measured in the laboratory.

\subsection{The role of standard stars}

Radial-velocity standards, with a supposedly known `true' velocity,
have long been used as objects
against which to calibrate observations of other stars.  Indeed,
a goal of the IAU Working Group on Radial Velocity Standard Stars
has been `to provide a list of such standard stars whose velocities
are known with an accuracy of 100~m~s$^{-1}$.'  Overviews of the
work by this and other groups are in \citet{stef+99,udry+99a,feke99},
and in various triennial reports from IAU Commission 30 on radial
velocities, published in the IAU Transactions
\citep[e.g.][]{berg94,ande99,rick01}.

In practical application, however, a number of dependences on the
0.5~km~s$^{-1}$ level have been found, in particular
on the stellar colour index, differences among different
radial-velocity instruments, between different spectrum correlation
masks applied on the same stellar spectrum, etc.  The best agreement
is normally found for stars of spectral type close to that of the Sun --
naturally so, since the instruments and data reductions are normally
calibrated against the solar spectrum e.g.\ as reflected off minor
planets, whose motions are accurately known through other methods,
and the procedures set up to produce consistent results at least for
such a solar spectrum.

However, not even very elaborate calibrations are likely to produce
any `true' standards to a much better precision than 0.5 (or,
perhaps, 0.3)~km~s$^{-1}$, unless a detailed physical model of the
observed star is developed.  The reason is simply the physical
nature of stellar spectra and the practical impossibility to
obtain noise-free measurements.  Apart from the physical effects
of stellar surface convection and gravitational redshifts mentioned
above, the wavelengths of stellar spectral lines depend on, i.a.,
factors such as the stellar rotation rate, the angle under which
the stellar rotation axis is observed, the phase in a possible
magnetic stellar activity cycle, and the spectral resolution and
instrumental profile of the observer's instrument.  Here, we give
examples of such effects that are likely to limit the ultimate
achievable precision for radial-velocity standard-stars to levels
not much better than 0.5~km~s$^{-1}$:

\subsubsection{Effects of stellar rotation}

The influence of stellar rotation has been realised, especially for
earlier-type stars with their often rapid rotation
\citep{ande+83,vers+99a,grif+00}.
The effects caused by the mismatch between a spectrum template for
a slow-rotation star and a rapidly rotating A-type star may exceed
1~km~s$^{-1}$ \citep{vers+99a}.  The origin of these effects is the
rotational line-broadening and the ensuing blending of spectral
lines, significant when a given star is
observed equator-on, but disappearing when viewed pole-on.

Even very modest rotational velocities in sharp-lined late-type
stars may cause significant wavelength displacements of the
spectral-line bottoms and other parts of the line profiles used
for radial-velocity determinations.  Naively, one might expect that
increased stellar rotation would merely smear out the line profiles
and perhaps straighten out the bisectors which describe the line
asymmetry.  Actually, for more
rapid rotation, when the asymmetric line components originating
near the stellar limbs begin to affect the wings of the profile
integrated over the stellar disk, the line asymmetries may become
enhanced.  This phenomenon was suggested by
\citet{gray+85} and by \citet{gray86}
and studied in more detail for a simulated rapidly rotating Sun by
\citet{smit+87}.  Detailed line profile calculations from
hydrodynamic model atmospheres for stars of different types
\citep{drav+90} show that this effect may easily displace
photospheric line-bottoms by a few hundred m~s$^{-1}$ already
in solar-type stars rotating with $V\sin i$ less than 10~km~s$^{-1}$.
Also, stars may rotate not as rigid bodies, and rotation may be
differential with respect to stellar latitude, atmospheric height,
or between magnetic and non-magnetic elements.
The existence of differential rotation is suggested both from studies of
starspots \citep[e.g.,][]{webe+01,coll+02},
and from analyses of line profiles \citep{rein+01}.

\subsubsection{Effects of stellar activity cycles}

On levels of perhaps 10--100~m~s$^{-1}$, at least cooler 
stars undergo apparent radial-velocity variations during a stellar 
activity cycle, when differently large fractions of the stellar 
surface are covered by active regions with magnetically `disturbed'
granulation \citep[e.g.,][]{gray+96}.
Magnetic flux that becomes entangled among the
convective features limits the sizes and the temperature and
velocity amplitudes to which these features develop. For
solar observations, see \citet[][{}~their Fig.~1]{spru+90} and
\citet{schm+88}; for theory, see \citet{berc+98}.
The resulting spectral line asymmetries are changed in the sense
of smaller asymmetries and smaller wavelength shifts in the
active regions
\citep[e.g.][{}~and references therein]{imme+89,bran+90}.

\citet{livi+99} followed the full-disk asymmetries of \ion{Fe}{i}
lines during a full 11-year solar activity cycle, finding cyclic
variations in the line asymmetry with an amplitude of about
20~m~s$^{-1}$; presumably the corresponding absolute shifts are
(at least) of a similar size.  Indeed, variations on this order
can be predicted from spatially resolved observations of line
profiles in active regions, weighted with the cyclically changing
area coverage of active regions during an activity cycle.
The effects can be expected to increase (to perhaps 50~m~s$^{-1}$)
for younger and chromospherically more active stars, e.g., 
F- and G-type ones in the Hyades \citep{saar+97}.

Since the amount of convective lineshift differs among different types of
spectral lines and between different spectral regions, also the
activity-induced changes in this shift must be expected to differ.  While
the identification of such differences could be important to find lines
whose sensitivity to stellar activity is smaller (thus being better
diagnostics for exoplanet signatures) or greater (being better diagnostics
for magnetic activity), such data are not yet available (and may indeed
require a stringent definition of the radial-velocity measure to permit
intercomparisons between observations at different epochs).

Besides these cyclic changes, there are shorter-term fluctuations
(on a level of perhaps 20--30~m~s$^{-1}$) in the apparent radial
velocity of stars, which often are greater in stars with enhanced
chromospheric activity.  Presumably, this reflects the evolution
and changing area coverages of active regions
\citep[e.g.][]{saar+98,saar+00,sant+00}.

\subsubsection{Effects caused by starspots and surface inhomogeneities}

Greater effects are present in spotted stars whose photometric variability
indicates the presence of dark spots across the stellar surface.
The amplitude of variations expected from photometrically dark spots
is on the order of 5~m~s$^{-1}$ for solar-age G-type stars,
increasing to perhaps 30--50~m~s$^{-1}$ for younger and more active
stars \citep{saar+97,hatz02}.
In more heavily spotted stars (such already classically classified
as variables), the technique of Doppler imaging exploits the
variability of spectral-line profiles to reconstruct stellar surface
maps \citep[e.g.,][]{pisk+90},
but obviously any more accurate deduction of the radial motion
of the stellar centre-of-mass from the distorted spectral lines is
not a straightforward task.

Surface inhomogeneities causing such line distortions need not be
connected to photometrically dark (or bright) spots, but could be
chemical inhomogeneities across the stellar surface (with locally
different line-strengths) or just patches of granulation with
different structure \citep{tone+88}.

\subsubsection{Effects caused by the finite number of granules}

Even the spectrum of a hypothetical spot-less and non-rotating star
with precisely
known physical and chemical properties will probably still not be
sufficiently stable to serve as a `standard' on our
intended levels of accuracy.  One reason is the finite number of
convective features (granules) across the stellar
surface.  For the Sun, a granule diameter is on the order of
1000~km, and there exist, at any one time, on the order of $10^6$
such granules on the visible solar disk.  The spectrum of integrated
sunlight is made up as the sum of all these contributions: to make
an order-of-magnitude estimate, we note that each granule has a
typical velocity amplitude of 1--2~km~s$^{-1}$.  Assuming that they
all evolve at random, the apparent velocity amplitude in the average
will be this number divided by the square root of $10^6$,
or 1--2~m~s$^{-1}$.  This `astrophysical noise' caused by the
finite number of granules is a quantity that 
is becoming measurable also in solar-type stars in
the form of an excess of the power spectrum of spectral-line variability at
temporal frequencies of some mHz, corresponding to granular lifetimes on the
order of ten minutes \citep{kjel+99}.
Although it can also be modelled theoretically
\citep[e.g.][]{tram+98}, such modelling can only predict the power
spectrum and other statistical properties, not the instantaneous
state of any star.

The number of granules across the surfaces of stars of other spectral
types may be significantly smaller than for the Sun, and the
resulting `random' variability correspondingly higher.  For
supergiants, it has been suggested than only a very small number
of convective elements (perhaps only a few tens) coexist at any given
time, but even a star with $10^4$ granules would show
radial-velocity flickering an order of magnitude greater than
the Sun.

The effects are qualitatively similar for other types of
statistically stable structures across stellar surfaces,
e.g.\ p-mode oscillations, where various surface regions on the
star are moving with varying vertical velocities.  Since their
averaging across the star does not fully cancel, they are
detectable as lineshift variations on a level of several
m~s$^{-1}$ \citep[e.g.][]{hatz96,bedd+01,fran+02}, 
demonstrating another
application of precise radial-velocity measurements, as well as
the limitations to stellar wavelength stability.

\subsubsection{Effects caused by exoplanets}

Intrinsically stable stars may show variability on the
10--100~m~s$^{-1}$ level induced by orbiting exoplanets.
Of course, this is a true radial-velocity variation, but it
practically limits the selection of such stars as radial-velocity
`standards', since their use on the m~s$^{-1}$ level would require
detailed ephemerides for their various exoplanets. It can be
noted that 51~Peg used to be a radial-velocity standard star!

\subsubsection{Effects of instrumental resolution}

A different type of wavelength displacements is introduced by the
observing apparatus, in effect convolving the pristine stellar
spectrum with the spectrometer instrumental profile.  Since all
stellar spectral lines are asymmetric to some extent, their
convolution with even a perfectly symmetric instrumental profile
of an ideal instrument produces a different asymmetry, and a
different wavelength position e.g.\ of the line-bottoms.
Quantitative calculations
demonstrate how such effects reach 50~m~s$^{-1}$ and more, even
for high-resolution instruments \citep{bray+78,drav+90}.
Although this is `only' a practical limitation which, in
principle, could be corrected for if full information of the
instrumental response were available, this is not likely to be
possible in practice.  For example, differences of many tens of
m~s$^{-1}$ in measured lineshifts may result between spectrometers
with identical spectral resolutions (measured as full width at
half-maximum), but
which differ only in their amounts of diffuse scattered light
\citep{drav87}. Of course, even greater lineshifts could be
caused by asymmetric instrumental profiles.
Instrumental effects in spectroscopy are reviewed by \citet{drav94}
while methods for calibrating instrumental profiles are discussed by, 
e.g., \citet{vale+95} and \citet{endl+00}.

\subsubsection{Effects of instrumental design}

For observational modes not involving analyses of highly resolved
line profiles, but rather statistical functions such as the
cross-correlation between spectral templates, a series of other
instrumental effects may intermix with stellar ones. 
For example, if a detector/template combination predominantly measures a
signal from the blue spectral region, it may be expected to record a
somewhat greater spectral blueshift in cool stars since convective
blueshifts generally increase at shorter wavelengths (where a given
temperature contrast in surface convection causes a relatively greater
brightness contrast).  A red-sensitive system may give the opposite bias,
unless it is sensitive into the infrared, where the generally smaller
stellar atmospheric opacities make the deeper layers visible, with perhaps
greater convective amplitudes.

\subsubsection{Conclusions about spectroscopic `standard' stars}

Physical and instrumental effects, such as those listed above
(and others, such as errors in laboratory wavelengths), imply that
there most probably do not exist any stars whose spectral features
could serve as a real standard on precision levels better than perhaps
300~m~s$^{-1}$.
Of course, for poorer precisions -- perhaps around 0.5~km~s$^{-1}$
-- various standard sources, including the solar spectrum, may
continue to be used as before.  However, in order to deduce \emph{true}
velocities to high accuracy, all spectral observations -- of
`standard' stars and others -- must undergo a detailed physical
modelling of their emitted spectrum, and of its recording process.

\subsubsection{Possible future astrometric standard stars}

The recently realised accurate determination of stellar radial
motions through astrometric measurements opens the possibility of
having also radial-velocity standards determined independent of
spectroscopy.  While the ultimate limitations in thus obtainable
accuracies have not yet been explored (e.g., what is measured in
astrometry is the photocentre of the normally unresolved stellar
disk, whose coordinates may be displaced by starspots or other
features) there appear to be no known effects that in principle
would hinder such measurements to better than 100~m~s$^{-1}$
(or even 10~m~s$^{-1}$), at least for some nearby stars.  This
will require astrometric accuracies on the microarcsecond level,
and possibly extended periods of observations, but these
are expected to be reachable in future space
astrometry missions \citep{drav+99a}.

\subsection{Conclusions}

From the above discussion it is clear that numerous effects may
influence the precise amount of spectral-line displacements.
Among these, only local effects near the observer (i.e., within the
[inner] solar system) can be reliably calculated and compensated for.
In particular, these depend on the motion and gravitational
potential of the observer relative to the desired reference frame,
normally the solar-system barycentre.

Therefore, spectroscopic methods will not be able, in any
foreseeable future, to provide values of stellar radial motion
with `absolute' accuracies even approaching our aim of 1~m~s$^{-1}$.
Of course, this does not preclude that measurement precisions
(in the sense of their reproducibility) may be much better and
permit the detection of very small \emph{variations} in the radial
velocity of an object (whose exact amount of physical motion
will remain unknown) in the course of searching for, e.g.,
stellar oscillations or orbiting exoplanets.  In order to enable
further progress in the many fields of radial-velocity studies,
it seems however that more stringent accuracy targets have to be
defined, so that future observational and theoretical studies have
clear goals to aim at.

Astrometric radial velocities do not appear to have the same
types of limitations as those deduced from spectroscopic shifts,
and more lend themselves to definitions that can be transformed
to `absolute' physical velocities.  These, however, must be
stringently defined since different plausible definitions differ
by much more than our desired accuracies.

\section{Kinematic and astrometric radial velocity}
\label{sec:math}

We will now more thoroughly scrutinise the various geometric effects
entering the concept of `radial velocity', aiming at definitions that
are consistent at an accuracy level of 1~m~s$^{-1}$. This requires
first that a system of temporal and spatial coordinates is adopted;
then that the relevant parts of the astronomical event (Fig.~1) are
modelled in this system, consistent with general relativity at the
appropriate accuracy level; and finally that suitable conventions are
proposed for the parameterisation of the event.

The modelling of astrometric observations within a general-relativistic
framework has been treated in textbooks such as \citet{murr83},
\citet{soff89} and \citet{brum91}, and various aspects of it have been
dealt with in several papers
\citep[e.g.][]{stum85,back+86,klio+92,klio00a,klio01}.
Much of the mathematical development in this and the next Section is
directly based on these treatments, but adapted in order to present a
coherent background for the definition and explanation of the
radial-velocity concepts.

\subsection{Coordinate system (BCRS)}
\label{sec:bcrs}

Subsequently, $t$ and $\vec{r}\equiv(x,y,z)$ denote
coordinates in the Barycentric Celestial Reference System (BCRS)
adopted by the IAU 24th General Assembly \citep{rick01} and
discussed by \citet{brum+01}. The temporal coordinate
$t$ is known as the Barycentric Coordinate Time (TCB).
The BCRS is a well-defined relativistic 4-dimensional coordinate
system suitable for accurate modelling of motions and events
within the solar system. However, it also serves as a
quasi-Euclidean reference frame for the motions of nearby stars
and of more distant objects, thanks to some useful properties:
it is asymptotically flat (Euclidean) at great distances from
the Sun; the directions of its axes are fixed with respect to
very distant extragalactic objects; and the origin at the
solar-system barycentre provides a local frame in which nearby
(single) stars appear to be non-accelerated, as they experience
practically the same galactic acceleration as the solar system.
The axes are aligned with the celestial system of right ascension
and declination as realised, for example, by the Hipparcos and
Tycho Catalogues \citep{esa97}.

\subsection{The light-time equation}
\label{sec:lt}

As emphasised in Sect.~\ref{sec:geom}, the relation between the
events of light emission and light reception is fundamental for
describing the astronomical event resulting in a geometric or
spectroscopic observation. The two events are connected by the
light-time equation, from which the required spatial and temporal
transformations may be derived.

In the BCRS, let $\vec{r}_\ast(t)$
describe the motion of the star and $\vec{r}_{\rm obs}(t)$ that
of the observer. Assume that a light signal is emitted from the
star at time $t_\ast$, when the star is at the spatial coordinate
$\vec{r}_\ast \equiv \vec{r}_\ast(t_\ast)$ and its coordinate velocity
is $\vec{v}_\ast\equiv(\mbox{d}\vec{r}_\ast/\mbox{d}t)_{t=t_\ast}$.
Assume, furthermore, that the same light signal is received by
the observer at time $t_{\rm obs}$, when the observer is at the
spatial coordinate
$\vec{r}_{\rm obs} \equiv \vec{r}_{\rm obs}(t_{\rm obs})$ and
its coordinate velocity is
$\vec{v}_{\rm obs}\equiv(\mbox{d}\vec{r}_{\rm obs}/\mbox{d}t)_{t=t_{\rm obs}}$.
The light-time equation can now be written 
\begin{equation}\label{eq:lte}
t_{\rm obs} - t_\ast = c^{-1}\left| \vec{r}_\ast
- \vec{r}_{\rm obs} \right| + \gamma \left(
t_\ast,\vec{r}_\ast;t_{\rm obs},\vec{r}_{\rm obs} \right) \, .
\end{equation}
Here, $c=299\,792\,458$~m~s$^{-1}$ is the conventional
speed of light, $|\,|$ denotes the usual (Euclidean) vector norm,
and $\gamma$ is the relativistic delay of the signal along the
path of propagation from star to observer. The delay term is
required to take into account that the coordinate speed of light
in the presence of a gravitational field is less than $c$ in the
adopted metric, so that
the first term in Eq.~(\ref{eq:lte}) gives too small a value for the
light travel time. For present purposes it is sufficient to
describe the gravitational field by means of the total Newtonian
potential $\Phi(\vec{r})$ relative to the BCRS. For instance,
the gravitational field of the solar system is adequately
described by $\Phi(\vec{r})=G\sum_i M_i|\vec{r}-\vec{r}_i|^{-1}$,
where $G$ is the gravitational constant, and the sum is taken over
the different solar-system bodies having (point) masses $M_i$
located at coordinates $\vec{r}_i$.
To first order in $c^{-2}$, the coordinate speed of light in the
BCRS metric is given by
$|\mbox{d}\vec{r}/\mbox{d}t|=c(1-2\Phi/c^2)$. The time delay
per unit length is therefore $2\Phi/c^3$. Integrating this quantity
along the light path (which for this calculation can be taken
to be a straight line in the BCRS coordinates) gives%
\footnote{Since the perturbing bodies move during the light
propagation, $\vec{r}_i$ should be taken to be the position at the
time of closest approach of the photon to the perturbing body.
For a rigorous treatment, see \cite{kope+99a}.}
\begin{equation}\label{eq:dt}
  \gamma = \sum_i \frac{2GM_i}{c^3}\,\ln\left(\frac{\vec{k}'
  (\vec{r}_\ast-\vec{r}_i)+|\vec{r}_\ast-\vec{r}_i|}{\vec{k}'
  (\vec{r}_{\rm obs}-\vec{r}_i)+|\vec{r}_{\rm obs}-\vec{r}_i|}\right) \, .
\end{equation}
$\vec{k}$ is the coordinate direction from the observer to the star,
\begin{equation}\label{eq:k}
\vec{k}=\left|\vec{r}_\ast - \vec{r}_{\rm obs} \right|^{-1}
\left(\vec{r}_\ast - \vec{r}_{\rm obs} \right) \, .
\end{equation}

Subsequently, we shall mainly consider the gravitational field of the
Sun (index~$=\odot$), for which $2GM_\odot c^{-3}\simeq 9.85~\mu$s.
For objects as distant as the stars we can neglect $\vec{r}_\odot$
compared with $\vec{r}_\ast$ in the numerator of the argument to the
logarithm in Eq.~(\ref{eq:dt}).  Moreover, $\vec{k}$ is practically
parallel with $\vec{r}_\ast$, so the numerator becomes simply
$2r_\ast$, where $r_\ast = |\vec{r}_\ast|$.  (The error introduced
by this approximation is $<10^{-16}$~s for $r_\ast>1$~pc.)  The
denominator varies depending on the relative positions of the
Sun, observer and star, but is typically of the order of the
astronomical unit ($A$) for an observer on the Earth.
Thus $\gamma \sim (10~\mu\mbox{s})\ln(2r_\ast/A)$, or
$\sim 100~\mu$s for the nearest stars, $\sim 200~\mu$s at
$r_\ast=10$~kpc, and $\sim 300~\mu$s for objects at cosmological
distances.  This slowly varying delay of a few hundred microseconds
suffered by the light while propagating from the star to the
solar system is generally ignored.  Indeed, it would hardly make
sense to try to evaluate it, since the gravitational delays caused
by other bodies (in particular by the star itself) are not included.

However, there is also a rapidly varying part of $\gamma$ in
Eq.~(\ref{eq:dt}), caused by the motion of the observer with
respect to the Sun.
In order to separate the rapidly varying part of the delay
from the (uninteresting) long-range delay, we write
\begin{equation}\label{eq:dts}
\gamma = \gamma_\ast - \sum_i\frac{2GM_i}{c^3}\,\ln\!\left(\frac{
\vec{k}{'}(\vec{r}_{\rm obs}-\vec{r}_i)+
|\vec{r}_{\rm obs}-\vec{r}_i|}{A}\right)\! ,
\end{equation}
where
\begin{equation}\label{eq:g0}
\gamma_\ast = 2G\Big({\textstyle \sum_i M_i}\Big)
c^{-3}\,\ln\!\left(\frac{2r_\ast}{A}\right)
\end{equation}
is practically a constant for the star (cf.\ below).
Clearly any constant length could have served instead
of $A$ to separate the terms in Eq.~(\ref{eq:dts}).
Using the astronomical unit for this purpose is just an
arbitrary convention.

From the light-time equation we can now determine the
relation between the coordinate time interval $\mbox{d}t_\ast$
(e.g.\ representing one period of emitted radiation) and
$\mbox{d}t_{\rm obs}$ (the corresponding period of coordinate
time at the observer).  Writing $|\vec{r}_\ast-\vec{r}_{\rm obs}|
= \vec{k}{'}(\vec{r}_\ast-\vec{r}_{\rm obs})$ we have
\begin{equation}
\mbox{d}t_{\rm obs} - \mbox{d}t_\ast = \vec{k}{'}
(\mbox{d}\vec{r}_\ast - \mbox{d}\vec{r}_{\rm obs})c^{-1}
+ \mbox{d}\gamma \, ,
\end{equation}
from which
\begin{equation}
\frac{\mbox{d}t_\ast}{\mbox{d}t_{\rm obs}} =
\frac{1+\vec{k}{'}\vec{v}_{\rm obs}\, c^{-1}
-\mbox{d}\gamma/\mbox{d}t_{\rm obs}}
{1+\vec{k}{'}\vec{v}_\ast\, c^{-1}} \, .
\end{equation}
From Eqs.~(\ref{eq:dts})--(\ref{eq:g0}) it follows that
$\mbox{d}\gamma/\mbox{d}t_{\rm obs}$ is the sum of two terms, the
first of which $\simeq (10~\mu\mbox{s})\rho/r_\ast$, where
$\rho=\mbox{d}r_\ast/\mbox{d}t_{\rm B}$ is the astrometric
radial velocity of the star defined below (Sect.~\ref{sec:astr}).
For $|\rho|<c$ and $r_\ast>1$~pc this term is $< 10^{-13}$ and
therefore negligible.  The second term can be evaluated e.g.\ for
an observer in circular orbit (at 1~AU) around the Sun.  It reaches
a maximum value of $<10^{-9}$ when the observer is behind the Sun
so that the light ray from the object just grazes the solar limb.
The simple expression
\begin{equation}\label{eq:dtdt}
\frac{\mbox{d}t_\ast}{\mbox{d}t_{\rm obs}} =
\frac{1+\vec{k}{'}\vec{v}_{\rm obs}\, c^{-1}}
{1+\vec{k}{'}\vec{v}_\ast\, c^{-1}}
\end{equation}
is therefore always good enough to a relative accuracy better
than $10^{-9}$ ($<0.3$~m~s$^{-1}$ in velocity).

\subsection{Barycentric time of light arrival}
\label{sec:btoa}

Since distances to objects beyond the solar system are generally
not known very accurately, it would be inconvenient to use the
time coordinate $t_\ast$ for describing observations of phenomena
that occur at such great distances.  Instead, it is customary to
relate the observed events to the time scale of the observer.
For very accurate timing, e.g.\ as required in pulsar observations,
one must take into account both the geometrical (R{\o}mer)
delay associated with the observer's motion around the solar-system
barycentre, and the relativistic (Shapiro) delay caused by the
gravitational field of bodies in the solar system.

We define the \emph{barycentric time of light arrival} as
\begin{equation}\label{eq:tbdef}
t_{\rm B} = t_\ast + r_\ast\,c^{-1} + \gamma_\ast \, ,
\end{equation}
where $\gamma_\ast$ is given by Eq.~(\ref{eq:g0}).  That is,
$t_{\rm B}$ is the time of light emission delayed by the nominal
propagation time to the barycentre ($r_\ast/c$) plus that part of
the relativistic delay which is independent of the observer.
By means of Eqs.~(\ref{eq:dt}) and (\ref{eq:dts}) we
find that the barycentric time of light arrival becomes
\begin{eqnarray}\label{eq:tbobs}
t_{\rm B} &=& t_{\rm obs} + (r_\ast - R)\, c^{-1} \nonumber\\[3pt]
&& +\sum_i \frac{2GM_i}{c^3}\,\ln\left[\frac{
\vec{k}{'}(\vec{r}_{\rm obs}-\vec{r}_i)+
|\vec{r}_{\rm obs}-\vec{r}_i|}{A}\right] \, ,
\end{eqnarray}
where $R=|\vec{r}_\ast - \vec{r}_{\rm obs}|$ is the
topocentric coordinate distance to the star.  With
$\vec{r}_\ast = \vec{u}r_\ast$, where $\vec{u}$ is the
barycentric coordinate direction to the star, the R{\o}mer delay
can be expanded to give
\begin{equation}\label{eq:dtexp}
(r_\ast - R)\, c^{-1} \simeq
\frac{\vec{u}{'}\vec{r}_{\rm obs}}{c}
- \frac{|\vec{u}\times\vec{r}_{\rm obs}|^2}{2cr_\ast}
+ \frac{(\vec{u}{'}\vec{r}_{\rm obs})^3}{2cr_\ast^2}\, .
\end{equation}
For $r_\ast>1$~pc and $r_{\rm obs}\sim 1$~AU the maximum
amplitudes of the three terms are $\simeq 500$~s, $1$~ms,
and $4$~ns, respectively; neglected terms are of order
$<10^{-13}$~s.

While Eq.~(\ref{eq:tbdef}) formally \emph{defines} the
barycentric time of light arrival, it is clear that
Eqs.~(\ref{eq:tbobs})--(\ref{eq:dtexp}) must in practice
be used to calculate
it for any given observation.  In principle this also
requires that the distance $r_\ast$ is known, but only to
a moderate precision.  In many practical situations, the
curvature terms in Eq.~(\ref{eq:dtexp}) [depending on
$r_\ast^{-1}$ and $r_\ast^{-2}$] can be neglected.

\subsection{Definition of kinematic parameters}
\label{sec:kin}

Within the BCRS we define the kinematic parameters of a
star to be its coordinate $\vec{r}_\ast$ at time
$t_\ast$, and the instantaneous coordinate velocity at
the same instant,
$v_\ast=\mbox{d}\vec{r}_\ast/\mbox{d}t|_{t=t_\ast}$.
In a time interval around $t_\ast$ we have
\begin{equation}
\vec{r}_\ast(t) = \vec{r}_\ast(t_\ast) + (t-t_\ast)\vec{v}_\ast
+ {\cal O}(t-t_\ast)^2 \, .
\end{equation}
The six components of the phase-space vector
$(\vec{r}_\ast,\vec{v}_\ast)$
are the relevant elements for studies of galactic kinematics
and dynamics, for example integration of galactic orbits
(after transformation to a galactocentric system).

We define the \emph{kinematic radial velocity} as the component
of $\vec{v}_\ast$ along the barycentric direction
$\vec{u}=\vec{r}_\ast r_\ast^{-1}$:
\begin{equation}
v_r = \vec{u} {'} \vec{v}_\ast \, .
\end{equation}
The perpendicular component of the coordinate velocity is the
kinematic tangential velocity,
$\vec{v}_t=\vec{v}_\ast-\vec{u}v_r$.  The kinematic radial and
tangential velocities are equivalent to the `true' radial and
tangential velocities introduced by \citet{klio00a}.

\subsection{Definition of astrometric parameters}
\label{sec:astr}

The six components of $(\vec{r}_\ast,\vec{v}_\ast)$ are not
directly observable but can in principle be derived from
astrometric observations of the star.  There is consequently
an equivalent set of six \emph{astrometric parameters}, which
we now define.  Following the discussion in Sect.~\ref{sec:btoa},
the astrometric parameters are considered as functions of
$t_{\rm B}$.  Writing the barycentric coordinate of the star
as
\begin{equation}\label{eq:rast}
\vec{r}_\ast = \vec{u}r_\ast
\end{equation}
we define the celestial coordinates $(\alpha,\delta)$ of the star
at the epoch $t_{\rm B}$ by means of the components of the unit
vector $\vec{u}$ in the BCRS.  This gives the first two astrometric
parameters.  The third one is parallax, which we define as
\begin{equation}\label{eq:pi}
\pi = \frac{A}{r_\ast} \,
\end{equation}
\citep[cf.][]{klio00a}.
The rate of change of the barycentric direction is the
proper-motion vector,
\begin{equation}\label{eq:mu}
\vec{\mu} = \frac{\mbox{d}\vec{u}}{\mbox{d}t_{\rm B}} \, ,
\end{equation}
from which the proper motion components
$\mu_{\alpha*}=\dot{\alpha}\cos\delta$ and
$\mu_{\delta}=\dot{\delta}$ follow (the dot signifies
differentiation with respect to $t_{\rm B}$).
These five parameters are practically identical to the standard
astrometric parameters used, for instance, in the Hipparcos
Catalogue.%
\footnote{There is however a subtle difference, in that proper
motions in the Hipparcos Catalogue are formally defined as
$\mbox{d}\vec{u}/\mbox{d}T_{\rm B}$, where $T_{\rm B}$ is the
barycentric time of light arrival expressed on the Terrestrial Time
(TT) scale. The TT scale is essentially the observer's proper time,
and differs from the coordinate time (TCB) used in Eq.~(\ref{eq:mu})
by the average factor $\langle\mbox{d}T/\mbox{d}t\rangle\simeq
1-1.55\times 10^{-8}$ \citep{irwi+99}
[cf.\ Eqs.~(\ref{eq:dtt}) and (\ref{eq:zbapprox})].
The Hipparcos proper motions should
therefore be multiplied by 0.9999999845 in order to agree with
the present definition.}
The sixth astrometric parameter is the rate of
change in barycentric coordinate distance, which we call the
\emph{astrometric radial velocity}:
\begin{equation}
\rho = \frac{\mbox{d}r_\ast}{\mbox{d}t_{\rm B}} \, .
\end{equation}
The term is motivated because of the exact analogy with the
definition of the (astrometric) proper motion in Eq.~(\ref{eq:mu}).
Observationally, the astrometric radial velocity can in principle
be determined e.g.\ from the secular change in parallax,
$\rho = \Delta(A/\pi)/\Delta t_{\rm B}$ \citep{drav+99a}.
The vector $\vec{\mu}r_\ast$ may be called the astrometric
tangential velocity.  The astrometric radial and tangential
velocities are equivalent to the `apparent' radial and
tangential velocities introduced by \citet{klio00a}.

It should be noted that the astrometric radial velocity is conceptually
quite different from the spectroscopic radial-velocity measure to be defined
in Sect.~\ref{sec:rvm}.  The astrometric radial velocity refers to the
variation of the coordinates of the source, and therefore depends on
the chosen coordinate system and time scale.  By contrast, the outcome
of a spectroscopic observation is a directly measurable quantity and
therefore independent of coordinate systems.

\subsection{Transformation between kinematic and astrometric
parameters}
\label{sec:trans}

The barycentric coordinate of the star is immediately derived
from the astrometric parameters by means of
Eqs.~(\ref{eq:rast})--(\ref{eq:pi}), viz.:
\begin{equation}\label{eq:r}
\vec{r}_\ast=\vec{u}(A/\pi) \, .
\end{equation}
Taking the derivative of Eq.~(\ref{eq:tbdef}) with respect to
$t_{\rm B}$ gives
\begin{equation}
1 = \frac{\mbox{d}t_\ast}{\mbox{d}t_{\rm B}} + \frac{\rho}{c}
\left(1+\frac{2G\sum_i M_i}{c^2r_\ast}\right) \, .
\end{equation}
The second term between parentheses is $<10^{-13}$ for
$r_\ast>1$~pc.  To sufficient accuracy we have therefore
\begin{equation}
\frac{\mbox{d}t_\ast}{\mbox{d}t_{\rm B}} = 1 - \frac{\rho}{c} \, .
\end{equation}
If Eq.~(\ref{eq:rast}) is differentiated with respect to $t_\ast$
we find
\begin{eqnarray}\label{eq:vast}
\vec{v}_\ast &=& \frac{\mbox{d}\vec{u}}{\mbox{d}t_\ast}r_\ast
+\vec{u}\frac{\mbox{d}r_\ast}{\mbox{d}t_\ast} =
\left( \frac{\mbox{d}\vec{u}}{\mbox{d}t_{\rm B}}r_\ast
+\vec{u}\frac{\mbox{d}r_\ast}{\mbox{d}t_{\rm B}} \right)
\frac{\mbox{d}t_{\rm B}}{\mbox{d}t_\ast} \nonumber\\
&=& \left(\vec{\mu}r_\ast + \vec{u}\rho\right)
(1-\rho/c)^{-1} \, .
\end{eqnarray}
Separating the radial and tangential components we have
\begin{equation}\label{eq:rho2vr}
 v_r = \rho\,(1-\rho/c)^{-1}
\end{equation}
and
\begin{equation}\label{eq:mu2vt}
 \vec{v}_t = \vec{\mu}r_\ast\,(1-\rho/c)^{-1} \, .
\end{equation}
Equations~(\ref{eq:r}) and (\ref{eq:vast})--(\ref{eq:mu2vt}) provide
the complete transformation from astrometric to kinematic parameters.
For the inverse transformation, we immediately
obtain $\vec{u}$ and $\pi$ from $\vec{r}_\ast$ by means of
Eqs.~(\ref{eq:rast}) and (\ref{eq:pi}).  Multiplying
Eq.~(\ref{eq:vast}) scalarly with $\vec{u}$ gives
\begin{equation}\label{eq:vr2rho}
\rho = \frac{\vec{u}{'}\vec{v}_\ast}
{1+\vec{u}{'}\vec{v}_\ast\,c^{-1}} = v_r(1+v_r/c)^{-1} \, ,
\end{equation}
from which finally
\begin{equation}\label{eq:muast}
\vec{\mu} = \frac{\pi}{A}\;\frac{\vec{v}_\ast-\vec{u}
\vec{u}{'}\vec{v}_\ast}{1+\vec{u}{'}\vec{v}_\ast\,c^{-1}}
= (\pi/A)\,\vec{v}_t\,(1+v_r/c)^{-1} \, .
\end{equation}
The kinematic quantities $v_\ast$, $v_r$ and $v_t$
are coordinate speeds of the object and therefore physically bounded by the
local coordinate speed of light, which in the BCRS far away from the Sun
is very close to $c$. The astrometric radial velocity $\rho$ and the
astrometric tangential velocity $\vec{\mu}r_\ast$, on the other hand, are
apparent quantities which may numerically exceed the speed of light. This
is so because the denominator $1+v_r/c$ in Eqs.~(\ref{eq:vr2rho}) and
(\ref{eq:muast}) can become arbitrarily small for an object moving at great
speed towards the observer. Thus, $\rho<-c$ if $v_r<-\frac{1}{2}c$,
while $|\vec{\mu}r_\ast|>c$ if $v_t>c+v_r$. The effect is equivalent to
the standard kinematic explanation of the superluminal expansion observed
in many extragalactic radio sources \citep{blan+77,verm+94}.

\section{The spectroscopic parameter:
Barycentric radial-velocity measure}
\label{sec:rvm}

We have found that the naive
notion of radial velocity as the line-of-sight component of the
stellar velocity is ambiguous already in a classical
(non-relativistic) formulation.  In a relativistic framework the
observed shift depends on additional factors, such as the
transverse velocity and gravitational potential of the source and,
ultimately, the cosmological redshift.  Since these factors are
generally not (accurately) known to the spectroscopic observer,
it is impossible to convert the observed shift $z$ into a precise
kinematic quantity.

What \emph{can} be derived from spectroscopic radial-velocity
measurements is the wavelength shift $z_{\rm B}$ corrected for
the local effects caused by the motion of the observer and the
potential field in which the observation was made.%
\footnote{The index ${\scriptstyle\rm B}$ for barycentric signifies
that -- in contrast to the case in cosmology -- the shift (and
velocity $cz_{\rm B}$) is referred to the solar-system barycentre,
not the rest-frame defined by the cosmological microwave background.}
For convenience, this shift
may be expressed in velocity units as $cz_{\rm B}$, where $c$ is
the conventional value for the speed of light.  Although this
quantity approximately corresponds to radial velocity, its precise
interpretation is model dependent and one should therefore avoid
calling it `radial velocity'. The term \emph{radial-velocity measure}
was proposed by \citet{lind+99}, and accepted in the later IAU
resolution, emphasising both its connection with the traditional
spectroscopic method and the fact that it is not quite the radial
velocity in the classical sense.

\subsection{The observed spectral shift}

In the previous sections the time coordinates of the various
events were all expressed on a single time scale $t$, i.e.\ the
Barycentric Coordinate Time (TCB).  As we now move on to
consider spectroscopic measurements, it is necessary to
include proper time ($\tau$) in our discussion.  The reason
for this is that the atomic transitions generating spectral 
lines can be regarded as oscillators or clocks that keep local
proper time. The measurement of spectroscopic line shifts is
therefore equivalent to comparing, by means of light signals,
the apparent rates of two identical atomic clocks, one located
at the source and the other at the observer.

Let $\tau_\ast$ be the proper time at the source of radiation,
and $\tau_{\rm obs}$ the proper time of the observer.  Suppose
that $n = \nu_\ast\mbox{d}\tau_\ast$ cycles of radiation are
emitted at frequency $\nu_\ast$ in the interval $\mbox{d}\tau_\ast$
of proper time at the source.  Let us also suppose that the
$n$ cycles are received in the interval $\mbox{d}\tau_{\rm obs}$
of proper time of the observer, who consequently derives the
frequency $\nu_{\rm obs}=n/\mbox{d}\tau_{\rm obs} =
\nu_\ast\mbox{d}\tau_\ast/\mbox{d}\tau_{\rm obs}$.  In terms
of wavelength ($\lambda=c/\nu$) the observed spectroscopic
shift $z_{\rm obs}$ is
\begin{equation}\label{eq:zobs}
1+z_{\rm obs} \equiv \frac{\lambda_{\rm obs}}{\lambda_\ast}
= \frac{\nu_\ast}{\nu_{\rm obs}}
= \frac{\mbox{d}\tau_{\rm obs}}{\mbox{d}\tau_\ast} \, ,
\end{equation}
where $\lambda_\ast$ ($=\lambda_{\rm lab}$) is the rest-frame
wavelength of the spectral line.  A spectroscopic lineshift
measurement is therefore equivalent to a direct comparison of
the proper time scales at the source and observer.  We
need to relate these proper time scales to the coordinate time
$t$ used in previous sections.

The relation between proper time and coordinate time is
defined by the adopted metric.  For the Barycentric
Celestial Reference System the accurate transformation can
be found for instance in \citet{peti00}.  For the present
applications we can ignore terms of order $c^{-4}$, leading
to the simple transformation
\begin{equation}\label{eq:dtt}
\frac{\mbox{d}\tau}{\mbox{d}t}(t,\vec{r},\vec{v})
= 1 - \frac{1}{c^2}\left(\Phi(\vec{r})
+ \frac{|\vec{v}|^2}{2}\right) \, ,
\end{equation}
in which $\Phi(\vec{r})$ is the Newtonian potential
introduced in Sect.~\ref{sec:lt} and
$\vec{v}=\mbox{d}\vec{r}/\mbox{d}t$ the
coordinate velocity.  This transformation
applies both to the source ($\mbox{d}\tau_\ast/\mbox{d}t_\ast$)
and the observer ($\mbox{d}\tau_{\rm obs}/\mbox{d}t_{\rm obs}$).
Thus, using Eq.~(\ref{eq:dtdt}), we have
\begin{eqnarray}\label{eq:dtaudt}
\frac{\mbox{d}\tau_{\rm obs}}{\mbox{d}\tau_\ast}
&=& \frac{\mbox{d}\tau_{\rm obs}}{\mbox{d}t_{\rm obs}} \,
\frac{\mbox{d}t_{\rm obs}}{\mbox{d}t_\ast} \,
\frac{\mbox{d}t_\ast}{\mbox{d}\tau_\ast} \nonumber\\
&=& \left(1 - \frac{\textstyle\Phi_{\rm obs}}{\textstyle c^2}
- \frac{\textstyle |\vec{v}_{\rm obs}|^2}{\textstyle 2c^2}\right)
\times
\left(1+\frac{\textstyle\vec{k}{'}\vec{v}_{\rm obs}}
{\textstyle c}\right)^{-1} \nonumber\\
&& \times \left(1+\frac{\textstyle\vec{k}{'}\vec{v}_\ast}
{\textstyle c}\right) \times
\left(1 - \frac{\textstyle \Phi_\ast}{\textstyle c^2}
- \frac{\textstyle |\vec{v}_\ast|^2}{\textstyle 2c^2}\right)^{-1} \, .
\end{eqnarray}
According to Eq.~(\ref{eq:zobs}) this equals the observed
wavelength ratio $1+z_{\rm obs}$.

\subsection{Reduction to the barycentre}

The first two factors on the right-hand side of Eq.~(\ref{eq:dtaudt})
depend on local conditions such as the motion of the observer in
the BCRS and the gravitational potential of the observer.  These
vary between different times and locations of
observers in the solar system, but they are also computable
to high accuracy from known data, including the barycentric position
and velocity of the observer.  The last two factors, on the other
hand, contain several quantities that cannot be uniquely separated
based on spectroscopic observations.  They depend on the
line-of-sight component of the star's coordinate velocity
($\vec{k}{'}\vec{v}_\ast$), but also on the gravitational potential
in the light-emitting region and (through $|\vec{v}_\ast|^2$) on
the tangential velocity of the star.

Let $z_{\rm B}$ be the spectral shift corrected for the local,
accurately computable effects, i.e.\ reduced to the solar-system
barycentre. In the approximation of
Eq.~(\ref{eq:dtaudt}) we have:
\begin{eqnarray}\label{eq:zb}
1+z_{\rm B} &=& (1+z_{\rm obs})
\left(1 - \frac{\textstyle\Phi_{\rm obs}}{\textstyle c^2}
- \frac{\textstyle |\vec{v}_{\rm obs}|^2}{\textstyle 2c^2}\right)^{-1}
\nonumber\\ && \times
\left(1+\frac{\textstyle\vec{k}{'}\vec{v}_{\rm obs}}
{\textstyle c}\right) \, .
\end{eqnarray}
It is important to note that the unit vector $\vec{k}$ in
Eq.~(\ref{eq:zb}) is the \emph{coordinate direction} to the star
given by Eq.~(\ref{eq:k}), not the observed (aberrated and refracted)
direction.

We now define \emph{barycentric radial-velocity measure} as the
quantity $cz_{\rm B}$, where $c=299\,792\,458$~m~s$^{-1}$.
For convenience, $cz_{\rm B}$ is expressed in velocity units through
multiplication with the constant $c$. The radial-velocity measure
therefore obtains physical dimensions of SI metres per SI second.%
\footnote{Naturally, $cz_{\rm B}$ can be expressed in m~s$^{-1}$ or
km~s$^{-1}$ according to convenience, and for cosmological velocities
the dimensionless measure $z_{\rm B}$ may be preferred.}
The epoch of any spectroscopic
observation should be given as the corresponding
barycentric time of light arrival (Sect.~\ref{sec:btoa}).

For an observer on the surface of the Earth (index $=\oplus$) we have
$\langle\Phi_{\rm obs}\rangle\simeq GM_\odot/A+GM_\oplus/R_\oplus\simeq
8.934\times 10^8$~m$^2$~s$^{-2}$ and
$\langle v_{\rm obs}\rangle\simeq 29785$~m~s$^{-1}$,
so that the second factor on the right-hand side of Eq.~(\ref{eq:zb})
is on the average
\begin{equation}\label{eq:zbapprox}
\left(1 - \frac{\textstyle\Phi_{\rm obs}}{\textstyle c^2}
- \frac{\textstyle |\vec{v}_{\rm obs}|^2}{\textstyle 2c^2}\right)^{-1}
\simeq 1+1.550\times 10^{-8} \, .
\end{equation}
In velocity units, the respective contributions from the solar
gravitational potential, the Earth's own gravitational potential,
and the Earth's velocity correspond to 3.0, 0.2 and 1.5~m~s$^{-1}$.
The main variation in this factor comes from the annual variation in
the observer's speed and distance from the Sun due to the eccentricity
($e\simeq 0.01671$) of the Earth's orbit. The resulting amplitude is
$2eGM_\odot/c^2\simeq 3.3\times 10^{-10}$, or 0.1~m~s$^{-1}$ in velocity
units. \emph{For an Earth-bound observer}, therefore, we may to
sufficient accuracy ($\sim$\,0.1~m~s$^{-1}$) use the average factor
in Eq.~(\ref{eq:zbapprox}) when reducing the observed shift to the
barycentre.

\subsection{Interpretation of the radial-velocity measure}
\label{sec:inter}

To first order, $cz_{\rm B}$ corresponds to the `classical'
spectroscopic radial velocity.  However, we emphasise that the radial-velocity
measure is just a quantification of the spectroscopic
shift, not of physical velocity.  Indeed, the interpretation of
$cz_{\rm B}$ in terms of a kinematic or astrometric radial velocity
is non-trivial and perhaps even impossible at the desired accuracy
level.  This is compounded by the additional effects discussed in
Sect.~\ref{sec:spec}, e.g.\ from motions in the stellar
atmosphere, pressure shifts, and cosmological redshift.  These
effects were ignored in Eq.~(\ref{eq:dtaudt}) and we now introduce
an extra factor $1+X$ to take them into account.  The
barycentric radial-velocity measure is then given by
\begin{equation}\label{eq:zb1}
1+z_{\rm B} = \left(1+\frac{\textstyle\vec{k}{'}\vec{v}_\ast}
{\textstyle c}\right)
\left(1 - \frac{\textstyle \Phi_\ast}{\textstyle c^2}
- \frac{\textstyle |\vec{v}_\ast|^2}{\textstyle 2c^2}\right)^{-1}
(1+X) \, .
\end{equation}
Only if $X$, $\Phi_\ast$ and the tangential velocity
$v_t$ are known to sufficient accuracy is it possible to derive
$\vec{k}{'}\vec{v}_\ast$ from the barycentric radial-velocity
measure.  Using also the distance information, the kinematic
radial velocity $v_r=\vec{u}{'}\vec{v}_\ast$ follows, and hence
the astrometric radial velocity from Eq.~(\ref{eq:vr2rho}).
Accurate transformation of $cz_{\rm B}$ to $v_r$ or $\rho$ is
therefore possible only in special circumstances.

\section{The IAU resolutions, and their application}
\label{sec:iau}

Based on the above discussion, and an interchange of opinions
in the community during a few years, two resolutions 
for the stringent definition of spectroscopic and astrometric
radial-velocity concepts were adopted at the IAU XXIVth General
Assembly held in Manchester, August 2000 \citep{rick02}.
Their full text is in the Appendix; in this Section we comment
on their practical implications.

\subsection{`Barycentric radial-velocity measure': A stringent
definition for spectroscopic measurements}

Briefly, the first resolution defines the \emph{barycentric
radial-velocity measure} $cz_{\rm B}$ as the result of a
spectroscopic measurement of line shifts; here $c$ is the speed
of light and $z_{\rm B}$ the wavelength shift referred
to the solar-system barycentre.  The definition avoids
any discussion on what the `true' radial velocity of the object
would be.  The transformation between the spectroscopically
determined barycentric radial-velocity measure $cz_{\rm B}$
and the physical velocity of the object is model-dependent and
cannot be treated in isolation from, e.g., the tangential motion
(cf.\ Sect.~\ref{sec:inter}).

The definition implies that high-accuracy radial-velocity
observations should be reduced to the solar-system barycentre
according to procedures based on general relativity and
using constants and ephemerides consistent with the required
accuracy.

\subsection{Practical application of the spectroscopic
definition}

For work at modest accuracies, the new definition implies no
change of existing procedures, nor of any published
radial-velocity values.  The use of the
`barycentric radial-velocity measure' will only be required
when absolute accuracies on the sub-km~s$^{-1}$ are needed.
However, its use permits to exploit radial-velocity
measurements for new classes of tasks, such as studying the
physical processes in stellar atmospheres exemplified in
Sect.~\ref{sec:spec}.

\subsubsection{Publishing observed wavelength shifts}

Traditionally, most published values for (stellar) radial
velocities have been transformed by the observer to some
`standard' system: instrumental; calibrated against standard
stars; against the spectrum of sunlight; or other.  The
traditionally reached precision has often been on the
order of 1~km~s$^{-1}$ or perhaps slightly better.  However,
given that the recently much improved measuring precisions
have begun to reach levels of m~s$^{-1}$, 
while absolute calibrations are only some order of magnitude worse, 
this procedure should change. The main point in the definition
of the `radial-velocity measure' is that highly precise
observations should be published (also) without the observer
trying to calibrate them against purported `standard'
objects in an effort to deduce the objects' physical
velocities.  Rather, the observations should be reduced
to the solar-system barycentre, as detailed in the
IAU resolution, and any subsequent interpretation of these
observed wavelength or frequency displacements in terms of
the object's motion, or other effects, should be made
separately.  The uncertainties in any attempted deduction
of the physical velocity are likely to be much greater
than those currently reachable in measurements of the
wavelength shifts themselves.  Therefore, any precise
observational data are likely to become corrupted by
applying such model-dependent `corrections', rendering
the data useless for possible more sophisticated analyses
in the future.

The `barycentric radial-velocity measure' is a quantity
that may be quite different for different spectral lines in
the same star, or for different portions of the same
spectral feature.  Nice examples of this are seen in
\citet{alle+02}, where weak absorption lines in the spectrum
of Procyon are observed to be systematically blueshifted
by almost 1000~m~s$^{-1}$ from the strong lines.  Further,
the wavelength positions of the line-bottoms are blueshifted
by some 200~m~s$^{-1}$ relative to those of the line flanks
closer to the continuum.  These particular effects can be
well modelled by hydrodynamical model atmospheres, and are
found to be caused by correlations between temperature and
vertical velocity in stellar surface convection.
Thus, precise radial-velocity measures
may be used as a novel tool to diagnose stellar hydrodynamics
\citep{drav02}, provided the data have not been corrupted by
futile attempts to `calibrate' the apparent velocities.

High-precision spectrometers often use some spectral template
with which the observed spectrum is cross correlated in order
to obtain a wavelength shift.  For any one stellar spectrum,
the resulting wavelength shift will naturally depend on the
exact properties of each different template (which portions
of what types of lines are being selected), in which
particular wavelength region are the measurements being made,
as well as on other parameters
\citep{grif+00,vers+99a,gull99b,gull+02}.
To retain the maximum amount of information and permit later
physical modelling, the barycentric radial-velocity measure
should be given together with details of the spectral
template and the correlation procedure.  Templates may be
constructed from both actual stellar spectra and lists of
laboratory wavelengths.  Since the former depend on spectrometer
resolution, and the latter are subject to revision as better
laboratory data become available, all such templates should
be fully documented, as should the software used for the
cross correlation (e.g., exactly what is being correlated:
the residual flux, or the line absorption; what is the
weighting of different spectral orders; exactly how are the
observed line shifts converted into velocity values?).

\citet{gull+02} deduced barycentric radial-velocity measures
for some forty stars (using only \ion{Fe}{i} lines)
with a median internal error of 27~m~s$^{-1}$, and an external
error of 120~m~s$^{-1}$ (the latter mainly coming from
uncertainties of the wavelength scale in the solar spectral
atlas used as wavelength reference).  Although the precision
achieved is somewhat lower than otherwise possible, the
\emph{accuracy} is higher since the procedures involved are fully
documented.
Such radial-velocity measures therefore become reproducible by 
other observers using different instruments and different techniques, 
as evidenced by the good agreement for those stars in common with 
\citet{nide+02}.

With improved measuring precisions, an increasing number of
publications have started to use expressions of `absolute'
velocities, often meaning merely the use of a zero-point on
the radial-velocity scale, obtained through calibrations
against the solar spectrum or otherwise.  The use of such
a term is somewhat unfortunate since the concept of absolute
velocity has a special physical meaning in relativity,
denoting something rather more fundamental than, e.g.,
certain modes of calibrating wavelength-shift measurements.

\subsubsection{Data reduction and software}

The velocity values obtained as a result from spectroscopic
observations depend not only on instrumental hardware effects,
but increasingly also on the software versions used for
reducing the data.

Detailed procedures for the reduction of spectroscopic
observations to the solar-system barycentre have been developed
e.g.\ by \citet{stum77,stum79,stum80b,stum85,stum86} and \citet{mcca95},
based on the series of solar-system ephemerides available from
JPL \citep{stan90}. The ephemerides and related computation services
are conveniently available on-line through JPL's HORIZONS System.%
\footnote{http://ssd.jpl.nasa.gov/}

At the 1994 IAU General Assembly it was decided to
systematically set up software tools in order to enhance the
interchangeability of observational data and theoretical ideas.
This set of tools is called the IAU Standards of Fundamental
Astronomy, SOFA \citep{fuku95,wall98}.
The SOFA collection of algorithms include such for the accurate
relativistic transformation of observed spectral-line
displacements to the solar-system barycentre.

Such algorithms apply to any periodic signals from a distant
source, not only the periodic modulation inherent to an
electromagnetic wave.  In particular, the periodic modulation
of pulsar signals follows the same mathematical physics.
In pulsar timing observations, the issues of uniform calibration
of observations from different stations, and their referral to
the solar-system barycentre, have been the topic of detailed
examinations \citep{hell86a,hell86b}. These transformations are
for instance included in software packages such as TEMPO, a
program for the analysis of pulsar timing data, maintained by
Princeton University and the Australia Telescope National Facility.%
\footnote{http://pulsar.princeton.edu/tempo/}
Many of these issues
(including those of defining reference frames, timescales, etc.)
are directly applicable also to electromagnetic waves.

The widespread FITS format
\citep[Flexible Image Transport System;][]{well+81}
used for the handling of astronomical data is undergoing
various modifications and extensions, including a more
elaborate representation of spectral quantities
\citep{grei+02}.
Alternative representations of spectral coordinates include
`radio-convention velocities' computed from frequency shifts,
`optical-convention velocities' computed from wavelength
shifts, `relativistic Doppler velocities', and others
(cf.\ Sect.~\ref{sec:dopp}).
While the differences among such concepts may be small for
most ordinary applications, any work aiming at very high
accuracy should carefully examine the exact definitions of
the various data fields, to understand how they can be
transformed to barycentric radial-velocity measures.
One has to remember that the prime purpose of standards
such as FITS is not the accurate physical interpretation
of data, but rather their transportation between different
computers and software environments.

\subsection{`Astrometric radial velocity': A stringent
definition for geometric measurements}

The second resolution simply specifies how `distance' and
`time' should be defined in order to provide the geometric
measurement of radial motion called \emph{astrometric radial
velocity}. Briefly, the resolution states that the appropriate
coordinate system is the Barycentric Celestial Coordinate System
(BCRS, Sect.~\ref{sec:bcrs}), with time expressed as the barycentric
time of light arrival (Sect.~\ref{sec:btoa}) on the barycentric
coordinate time scale (TCB).
Analogously, the conventional understanding of proper
motion is generally understood to mean the rate of change
in barycentric direction with respect to the barycentric
time of light arrival, although we are not aware of any previous
formal definition to that effect.

\section{Unsolved issues}
\label{sec:unsolved}

The IAU resolutions were elaborated with the aim to permit results
of spectroscopic and astrometric radial-velocity measurements to be
unambiguously quantified on the 1~m~s$^{-1}$ level. Many known effects
on the sub-m~s$^{-1}$ level are also taken care of in the present
definitions. For example, the radial velocity of any object varies
cyclically throughout the year, as the observer orbits the Sun and
views the stellar velocity vector under a slightly different projection
angle. Since the radial-velocity measure is defined relative to the
solar-system barycentre, such ambiguities are removed. However, there
do exist other issues, where the present concepts may be inadequate.
A few of them are highlighted below.

\subsection{Effects beyond the (inner) solar system}

The definition leaves `uncorrected' all the (largely unknown)
effects originating from outside the (inner) solar system.
In particular, the BCRS describes an asymptotically flat metric
at large distances from the Sun, thus ignoring effects of the
gravitational fields from other individual stars and, on a larger
scale, from the Milky Way Galaxy and structures therein (e.g.,
spiral arms, dark-matter concentrations). For instance, the
large-scale gravitational potential of the Galaxy causes wavelength
shifts that may be relevant for highly accurate kinematic modelling.
Over a range of galactocentric distances ($R$) the galactic potential
is crudely described by that of a singular isothermal sphere
\citep{binn+87}, leading to a differential gravitational redshift
between the star and observer of
$\Delta v_{\rm grav}=(V^2/c)\ln(R_{\rm obs}/R_\ast)$,
where $V\simeq 220$~km~s$^{-1}$ is the circular galactocentric speed.
Thus, the spectra of stars in the central bulge ($R_\ast \sim 1$~kpc)
may be gravitationally redshifted by 300--400~m~s$^{-1}$, while
stars in the Magellanic Clouds ($R_\ast \simeq 55$~kpc) might be
blueshifted by a similar amount, as seen by an observer 
near the solar position at $R_{\rm obs}\simeq 8.5$~kpc.

\subsection{Gravitational lensing}

Gravitational lensing, i.e., the bending or focusing of light
during its propagation through gravitational fields may affect
the radial velocity in different ways.  In gravitational microlensing,
another (fainter) star or other object passes very nearly in front of
the target star (as seen by the observer), and its gravitational
field focuses light toward the observer.  A stellar
gravitational field is too weak to cause resolvable multiple
images, so instead a source brightening is observed.
The velocities of stars in the Milky Way (acting as lenses)
imply typical timescales for such events on the order of a
few weeks ($\sim$\,$10^6$~seconds).

The gravitational field of the lens causes a time delay of
the light signal from the target star. This (Shapiro) delay,
given by Eq.~(\ref{eq:dt}), is of order $(20~\mu\mbox{s})\ln(r/p)$
for solar-mass lenses, where $r$ is the distance to the lens
(assumed to be half-way to the target) and $p$ the impact parameter
of the light ray. Thus for a ray grazing the stellar limb and observed
at 1~kpc distance the delay is of order 0.5~ms. A variation of the delay
by this amount over a timescale of $10^6$~s would cause an
apparent change in the radial-velocity measure of the target star by
$\sim$\,0.15~m~s$^{-1}$. Lensing by more massive or compact
objects could thus in principle produce measurable variations.
For a discussion of the corresponding effect on pulsar timing
observations, see \citet{hoso+99}.
Under certain conditions additional relativistic effects
causing time delays might enter, such as the Lense--Thirring
or Kerr delay caused by the spin of the gravitational
source and the ensuing frame-dragging.

However, quite different effects could cause much more significant
changes in the observed wavelengths of stellar spectral lines during
a microlensing event. The amount of light amplification from the
target object depends on the exact geometry of the target, the lens,
and the observer.  On this microarcsecond level, the disk of the
target star is an extended object, and different parts of its
disk gradually undergo different amounts of flux magnification,
as the lensing object passes by.  Since stars often rotate at
a significant rate, portions of the stellar disk that approach
the observer (with a spectrum Doppler-shifted to the blue)
may at some time be differently enhanced from the redshifted
portions near the opposite stellar limb (receding from the
observer), producing a variable wavelength shift on a level
of up to several km~s$^{-1}$ \citep{maoz+94,goul97}.

Gravitational lensing by more massive objects, e.g., clusters
of galaxies, often produces multiple or extended images of the
same target object.  Each (sub)image corresponds to a different
light-path to the source, and thus a different Shapiro delay.
For a geometry changing with time, there would be a variable
differential delay, causing each (sub)image
to have a different (spectroscopic) radial velocity. Thus a
particular source would not have one unique radial velocity,
but different values depending on which among several light-paths
from the source to the observer that are chosen.

\subsection{Gravitational waves}

The source of the gravitational lensing need not be stationary,
but could be transient, in the form of a passing gravitational
wave.  Although it appears that the effects will be very small,
there might exist specific conditions (such as compact objects
in close binary systems), where the variable time delays
introduced by such waves should be taken into consideration, at
least in principle; e.g., \citet{faki94,kope+99}.

\subsection{Cosmological effects}

A conceptual problem concerns the cosmological redshift: what is
the meaning of `radial velocity' in the context of an expanding
Universe? Is it to be understood as a motion relative to the
general expansion, as represented by the Hubble parameter
$H_0\simeq 70$~km~s$^{-1}$~Mpc$^{-1}=70$~m~s$^{-1}$~kpc$^{-1}$,
or relative to the local
expansion rate? Our accuracy aim of 1~m~s$^{-1}$ corresponds to
the formal expansion velocity at a distance of only 14~pc, that
of very nearby stars.

The extent to which local systems participate in the general
expansion of the Universe is a problem that has been treated by
several authors, beginning with \citet{mcvi33}.  It has been argued
that local entities such as the solar system or even the
Milky Way Galaxy should be unaffected by the cosmic expansion
since if everything expanded equally, the expansion would be
unobservable.  The full problem is quite complex,
but there seems to be no fundamental reason why there should
be a specific scale below which there is no expansion.
For detailed discussions, see, e.g.,
\citet{coop+98}, and references therein.

Finally, on cosmological scales of time and space, we cannot
even be certain about the constancy of physical `constants';
to include such possible effects in the definitions remains a
task for the future.

\begin{acknowledgements}
The initial concepts for the IAU resolutions were discussed
during a number of conferences in 1998--1999, and evolved in
e-mail discussions with hundreds of messages exchanged over
a couple of years; we thank the many colleagues that gave their
input.  Thanks are further extended to John Hearnshaw
(Christchurch, New Zealand) for organising and chairing the
multi-commission meeting at the IAU General Assembly in Manchester,
where these resolutions were adopted.  This work is supported
by the Swedish Research Council, and the Swedish National Space
Board.
\end{acknowledgements}

\appendix
\section*{Appendix: IAU resolutions}

To enable high-accuracy studies of radial velocities, and to
permit accurate comparisons between observers using different
methods, two resolutions were adopted by a number of Divisions
and Committees of the International Astronomical Union, at a
special session during its XXIVth General Assembly in Manchester
(August 2000).  The resolutions define a spectroscopic
\emph{barycentric radial-velocity measure}, and an \emph{astrometric
radial velocity}.  The full text of the resolutions follows
\citep{rick02}.%
\footnote{The texts of Resolutions C1 and C2 are also available at\\
http://www.astro.lu.se/$\sim$dainis/HTML/RADVEL.html and\\
http://www.astro.lu.se/$\sim$dainis/HTML/ASTRVEL.html\\
respectively. The resolutions referred to in the Note at the end of
Resolution C2 are found in IAU Transactions XXIV B, pp.~37--43
and 44--49, and at\\
http://danof.obspm.fr/IAU\_resolutions/Resol-UAI.htm}

\bigskip\bigskip\noindent
{\large\sffamily\itshape
Resolution C1 on the Definition of a Spectroscopic\\
"Barycentric Radial-Velocity Measure"}

\medskip\noindent
Divisions I, IV, V, VI, VII, IX and X, and Commissions 8, 27, 29,
30, 31, 33, 34 and 40 of the International Astronomical Union

\medskip\noindent
\emph{Recognising}

\begin{enumerate}
\item
That recently improved techniques for determining radial
velocities in stars and other objects, reaching and exceeding
precision levels of meters per second, require the definition
of "radial velocity" to be examined;
\item
That, due to relativistic effects, measurements being made
inside gravitational fields, and alternative choices of
coordinate frames, the naive concept of radial velocity being
equal to the time derivative of distance, becomes ambiguous at
accuracy levels around 100 m/s;
\end{enumerate}

\medskip\noindent
\emph{Considering}

\begin{enumerate}
\item
That, although many effects may influence the precise shifts
of spectroscopic wavelengths and frequencies, only local ones
(i.e. arising within the solar system, and depending on the
gravitational potential of the observer, and the observer's
position and motion relative to the solar-system barycenter)
can in general be reliably calculated;
\item
That, although the wavelength displacement (or frequency
shift) corrected for such local effects can thus be derived
from spectroscopic measurements, the resulting quantity cannot
unambiguously be interpreted as a radial motion of the object;
\end{enumerate}

\medskip\noindent
\emph{Therefore recommend}

\medskip\noindent
That, whenever radial velocities are considered to a high
accuracy, the spectroscopic result from a measurement of shifts
in wavelength or frequency be given as the "barycentric
radial-velocity measure" $cz_{\rm B}$, after correcting for
gravitational effects caused by solar-system objects, and effects by
the observer's displacement and motion relative to the solar-system
barycenter.

\medskip\noindent
Here, $c$ equals the conventional speed of light $=$ 299,792,458 m/s,
and $z_{\rm B} = (\lambda - \lambda_0)/\lambda_0$, where $\lambda_0$
is the rest-frame wavelength and $\lambda$ the wavelength observed
by a hypothetical observer at zero gravitational potential,
located at, and being at rest with respect to, the solar-system
barycenter.  The epoch of the observation equals the barycentric
time of light arrival.

\medskip\noindent
The radial-velocity measure $cz_{\rm B}$ is expressed in velocity
units: to first order in $z_{\rm B}$ it coincides with the classical
concept of "radial velocity", while avoiding the implicit
interpretation as physical motion.  The solar-system barycenter
is defined by Resolution A4 adopted at the IAU XXIst General
Assembly in 1991, and supplemented by Resolution B6 at the
IAU XXIIIrd General Assembly in 1997.

\bigskip\bigskip\noindent
{\large\sffamily\itshape
Resolution C2 on the Definition of "Astrometric\\ Radial Velocity"}

\medskip\noindent
Divisions I, IV, V, VI, VII, IX and X, and Commissions 8, 27, 29,
30, 31, 33, 34 and 40 of the International Astronomical Union

\medskip\noindent
\emph{Recognising}

\medskip\noindent
That recently improved astrometric techniques may permit the
accurate determination of stellar radial velocities independent
of spectroscopy, thus requiring a definition independent from
spectroscopic measures;

\medskip\noindent
\emph{Considering}

\medskip\noindent
That the change in the barycentric direction $\vec{u}$ to objects
outside of the solar system is customarily expressed by the
proper-motion vector $\vec{\mu}={\rm d}\vec{u}/{\rm d}t_{\rm B}$,
where $t_{\rm B}$ is the barycentric coordinate time (TCB)
of light arrival at the solar system barycenter;

\medskip\noindent
\emph{Therefore recommend}

\medskip\noindent
That the geometric concept of radial velocity be defined as
${\rm d}r/{\rm d}t_{\rm B}$, where $r$ is the barycentric
coordinate distance to the object and $t_{\rm B}$ the
barycentric coordinate time (TCB) for light arrival at the
solar system barycenter.

\medskip\noindent
Note: The Barycentric Celestial Reference System
(including the barycentric coordinate time) is defined in
Resolutions B1.3 and B1.5 adopted at the IAU XXIVth General
Assembly in 2000.

\bibliographystyle{aa}
\bibliography{h3961}

\end{document}